\documentclass[aps,showpacs,prb,twocolumn]{revtex4}
\usepackage{mathrsfs}
\usepackage{amssymb}
\usepackage{amsmath}
\usepackage{bm}
\usepackage{graphics}
\usepackage{natbib}

\begin{document}
\title{Three-terminal triple-quantum-dot ring as a charge and spin current rectifier}

\author{Weijiang Gong$^{1,2}$}
\author{Hui Li$^{1}$}
\author{Guozhu Wei$^{1,2}$}
\affiliation{ 1. College of Sciences, Northeastern University,
Shenyang 110819, China \\
2. International Centre for Materials Physics, Chinese Academy of
Sciences, Shenyang, 110016, China}
\date{\today}

\begin{abstract}
Electronic transport through a triple-quantum-dot ring with three
terminals is theoretically studied. By introducing local Rashba
spin-orbit interaction on an individual quantum dot, we find that
the spin bias in one terminal drives apparent charge currents in the
other terminals, accompanied by the similar amplitude and opposite
directions of them. Meanwhile, it shows that the characteristics of
the spin currents induced by the spin bias are notable. When a
magnetic flux is applied through this ring, we see its nontrivial
role in the manipulation of the charge and spin currents. With the
obtained results, we propose this structure to be a prototype of a
charge and spin current rectifier.

\end{abstract}
 \maketitle

\bigskip

\section{Introduction}
The manipulation and control of the behaviors of electron spins in
nanostructures have become one subject of intense investigation due
to its relevance to quantum computation and quantum
information.\cite{Ohno,Dasarma} The electron spin in quantum dot
(QD) is a natural candidate for the qubit, and then QD is regarded
as an elementary cell of quantum computation and quantum
information. Therefore, much attention has been paid to the
manipulation of the electron spin degree of freedom in QD for its
application.\cite{Loss,Loss2} Many schemes have been proposed to
work out this problem for obtaining the highly-polarized spin
current (in particular, the so-called \emph{pure spin current}),
based on the case of a charge bias between two leads with a magnetic
field or the spin-obit (SO) coupling for a QD
system.\cite{Datta,Rashba,Rashba2,Sun,Trocha,Other} Despite these
existed works, any new suggestions to realize the highly-polarized
spin current are still necessary. Recently, it has been reported
that spin bias in leads for mesoscopic systems can be feasible,
which, different from the traditional charge bias, induces rich
physical phenomena and potential
applications\cite{Nagaosa,Hubner,Cui,Li,Jpn,LuHZ,Chi}.
\par
In the present work, we choose a three-terminal triple-QD ring,
which is feasible to be fabricated by virtue of the current
nanoscale and mesoscale technology\cite{Nagaosa,Bao,Tarucha}, to
investigate its electron transport properties influenced by the spin
bias in one terminal. As a result, by introducing local Rashba
spin-orbit interaction on an individual QD, it is found that the
spin bias drives apparent charge currents in the two other terminals
with interesting properties of them. Besides, the characteristics of
the spin currents induced by the spin bias are also notable. With
these results, this structure can be proposed to be a prototype of a
charge and spin current rectifier.
\begin{figure}
\begin{center}
\scalebox{0.4}{\includegraphics{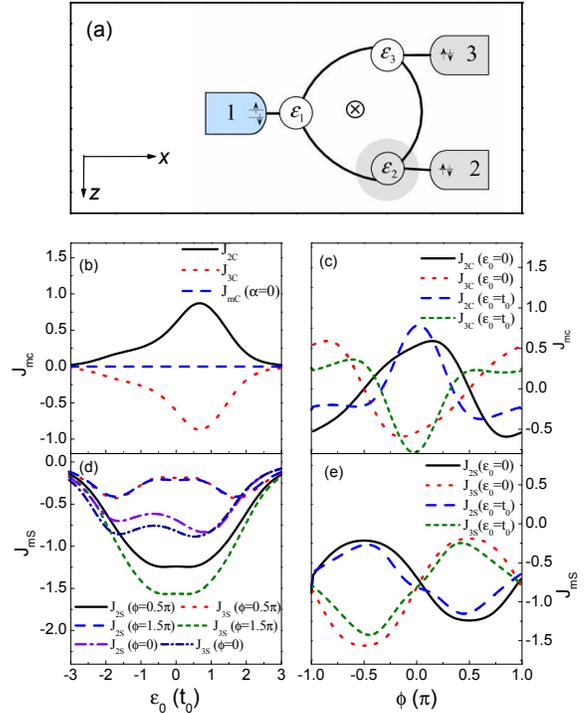}} \caption{ (a) Schematic
of a three-terminal triple-QD ring structure with a local Rashba
interaction on QD-2. Three QDs and the leads coupling to them are
denoted as QD-$j$ and lead-$j$ with $j=1-3$. Spin bias is assumed to
be in lead-1. (b) The charge current spectra in lead-2 and lead-3
with the shift of QD levels. (c)The charge currents vs the magnetic
phase factor $\phi$. (d) The spin current spectra in lead-2 and
lead-3. (e) The spin current spectra vs the magnetic phase factor
$\phi$. \label{Structure}}
\end{center}
\end{figure}
\section{model and formulation\label{theory}}
The considered three-terminal triple-QD ring is illustrated in
Fig.\ref{Structure}(a), which can usually be fabricated by means of
split-gate technique in the two-dimensional electron gas. We assume
that in one terminal ( lead-1 ) there exists the spin bias $V_{s}$,
i.e., the spin-dependent chemical potentials for the spin-up and
spin-down electrons are $\mu_{1\sigma}=\varepsilon_F+\sigma
eV_{s}$.\cite{Sun} By additionally inserting two normal metallic
terminals ( lead-2 and lead-3 ) ( here
$\mu_{2\sigma}=\mu_{3\sigma}=\varepsilon_F$ ), we would like to
observe the charge and spin transport behaviors in the other
terminals affected by the spin bias. Correspondingly, if lead-1 is
postulated to be the source terminal, the others will be the drain
terminals.
\par
In this structure, the Hamiltonian of the electron moving in the x-z
plane can be written as
$H_s=\frac{\textbf{P}^2}{2m^*}+V(\textbf{r})+\frac{\hat{y}}{2\hbar}\cdot[\alpha(\hat{\sigma}\times
\textbf{p})+(\hat{\sigma}\times \textbf{p})\alpha]$, where the
potential $V(\textbf{r})$ confines the electron to form the
structure geometry, namely, the leads, QDs and the connections. The
last term in $H_s$ denotes the local Rashba SO coupling on QD-2. For
the analysis of the electron properties, we have to second-quantize
the above Hamiltonian,\cite{gongapl} which is composed of three
parts: ${\cal H}={\cal H}_{c}+{\cal H}_{d}+{\cal H}_{t}$.
\begin{eqnarray}
{\cal H}_{c}&&=\underset{\sigma jk}{\sum }\varepsilon
_{jk\sigma}c_{jk\sigma}^\dag c_{jk\sigma },\notag\\
{\cal H}_d&&=\sum_{j=1, \sigma}^{3}\varepsilon
_{j}d^\dag_{j\sigma}d_{j\sigma}
+\sum_{l=1,\sigma}^{2}[t_{l\sigma}d^\dag_{l\sigma}d_{l+1\sigma}+r_l(d_{l\downarrow}^\dag
d_{l+1\uparrow}\notag\\&&-d_{l+1\downarrow}^\dag
d_{l\uparrow})]+t_{3}e^{i\phi}d^\dag_{3\sigma}d_{1\sigma}+\mathrm
{H.c.},\notag\\ {\cal H}_{t}&&=\underset{\sigma jk }{\sum
}V_{j\sigma} d^\dag_{j\sigma}c_{jk\sigma}+\mathrm {H.c.},
\end{eqnarray}
where $c_{jk\sigma}^\dag$ and $d^{\dag}_{j\sigma}$ $( c_{jk\sigma}$
and $d_{j\sigma})$ are the creation (annihilation) operators
corresponding to the basis in lead-$j$ and QD-$j$. $\varepsilon
_{jk\sigma}$ and $\varepsilon_{j}$ are the single-particle levels.
$V_{j\sigma}$ denotes QD-lead coupling strength. The interdot
hopping amplitude
$t_{l\sigma}=t_l\sqrt{1+\tilde{\alpha}^2}e^{-i\sigma\varphi}$
($l=1,2$), where $t_{l}$ is the ordinary transfer integral
irrelevant to the Rashba interaction and $\tilde{\alpha}$ is the
dimensionless Rashba coefficient with
$\varphi=\tan^{-1}\tilde{\alpha}$\cite{Serra}. $r_l$ is a complex
quantity representing the strength of interdot spin flip. The phase
factor $\phi$ attached to $t_{3}$ accounts for the magnetic flux
through the ring. In addition, the many-body effect can be readily
incorporated into the above Hamiltonian by adding the Hubbard term
${\cal
V}_{e\text{-}e}=\sum_{j\sigma}{\frac{U_j}{2}}n_{j\sigma}n_{j\bar{\sigma}}$.
\par
Starting from the second-quantized Hamiltonian, we can now formulate
the electronic transport properties. With the nonequilibrium Keldysh
Green function technique, the spin-$\sigma$ current flow in lead-$j$
can be written as\cite{Meir,Gong1}
\begin{equation}
J_{j\sigma}=\frac{e}{h}\sum_{j'\sigma'}\int d\omega
T_{j\sigma,j'\sigma'}(\omega)[f_{j\sigma}(\omega)-f_{j'\sigma'}(\omega)],\label{current}
\end{equation}
where $f_{j\sigma}(\omega)=(\exp{\omega-\mu_{j\sigma}\over
k_BT}+1)^{-1}$ is the Fermi distribution function in lead-$j$.
$T_{j\sigma,j'\sigma'}(\omega)=4\Gamma_ {j\sigma}
 G^r_{j\sigma,j'\sigma'}(\omega)\Gamma_{j'\sigma'}G^a_{j'\sigma',j\sigma}(\omega)$
is the transmission function, describing electron tunneling ability
between lead-$j$ to lead-$j'$. $\Gamma_{j\sigma}=\pi
|V_{j\sigma}|^2\rho_j(\omega)$, the coupling strength between QD-$j$
and lead-$j$, can be usually regarded as a constant.  $G^r$ and
$G^a$, the retarded and advanced Green functions, obey the
relationship $[G^r]=[G^a]^\dag$. From the equation-of-motion method,
the retarded Green function can be obtained in a matrix form,
\begin{eqnarray}
&&[G^r]^{-1}=\notag\\
&&\left[\begin{array}{cccccc} g_{1\uparrow}^{-1} & -t_{1\uparrow}&-t_{3}e^{-i\phi}&0&r^*_1&0\\
  -t^*_{1\uparrow}& g_{2\uparrow}^{-1}& -t_{2\uparrow}&-r^*_1&0&r^*_2\\
  -t_{3}e^{i\phi}&-t^*_{2\uparrow}&g_{3\uparrow}^{-1}&0&-r^*_2&0 \\
  0&-r_1&0&g_{1\downarrow}^{-1}&-t_{1\downarrow}&-t_{3}e^{-i\phi}\\
  r_1&0&-r_2&-t^*_{1\downarrow}& g_{2\downarrow}^{-1}& -t_{2\downarrow}\\
  0&r_2&0&-t_{3}e^{i\phi}&-t^*_{2\downarrow}&g_{3\downarrow}^{-1}
\end{array}\right]\notag.
\end{eqnarray}
In the above expression, $g_{j\sigma}$ is the Green function of
QD-$j$ unperturbed by the other QDs and in the absence of Rashba
effect.
$g_{j\sigma}=[(z-\varepsilon_{j})\lambda_{j\sigma}+i\Gamma_{j\sigma}]^{-1}$
with $z=\omega+i0^+$.
$\lambda_{j\sigma}=\frac{z-\varepsilon_{j}-U_{j}}{z-\varepsilon_{j}-U_{j}+U_j\langle
n_{j\bar{\sigma}}\rangle}$ results from the second-order
approximation of the Coulomb interaction\cite{ZhengHZ}, which is
reasonable when the system temperature is higher than the Kondo
temperature. $\langle n_{j\sigma}\rangle$ can be numerically
resolved by the formula $\langle
n_{j\sigma}\rangle=-\frac{i}{2\pi}\int d\omega
G^{<}_{j\sigma,j\sigma}$ where
$G^<_{\sigma\sigma}=\sum_{\sigma'}[G^r]_{\sigma\sigma'}[\Sigma^<]_{\sigma'}[G^a]_{\sigma'\sigma}$
and $[\Sigma^<]_\sigma=2\sum_j\Gamma_{j\sigma}f_{j\sigma}(\omega)$.
\begin{figure}
\begin{center}
\scalebox{0.4}{\includegraphics{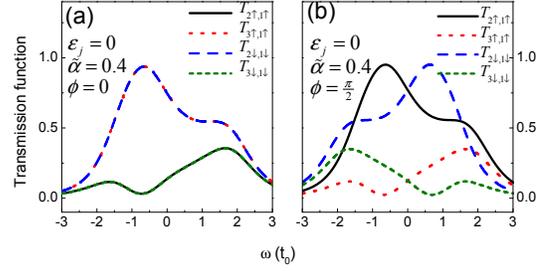}} \caption{ The spectra of
transmission functions $T_{m\sigma,1\sigma}$($m$=2,3) with the QD
levels fixed at $\varepsilon_j=0$. (a) and (b) Zero magnetic field
case, and (c)-(d) magnetic phase factor $\phi={\pi\over
2}$.\label{Trans}}
\end{center}
\end{figure}
\section{Numerical results and discussions \label{result2}}
\par
We now proceed on to calculate the charge currents in the drain
terminals, lead-2 and lead-3 in this case. Before calculation, the
QD-lead couplings are assumed to take the uniform values with
$\Gamma_{j\sigma}=t_0$, and we consider $t_0$ as the energy unit.
The structure parameters are for simplicity taken as
$|t_{l\sigma}|=t_3=t_0$, and $\varepsilon_F$ is viewed as the energy
zero point of this system. Besides, to carry out the numerical
calculation, we choose the Rashba coefficient $\tilde{\alpha}=0.4$
which is available in the current experiment.\cite{Sarra}
\par
We first focus on the electron transport in the linear regime. In
this case, the charge current flow is proportional to the linear
conductance, i.e., $J_{mc}=\mathcal {G}_{mc}\cdot V_s$ ($m=2,3$),
where $J_{mc}=J_{m\uparrow}+J_{m\downarrow}$ and the linear charge
conductance
\begin{eqnarray}
\mathcal {G}_{mc}=\frac{e^{2}}{h}\sum_\sigma[\bar{\sigma}
T_{m\sigma,1\sigma}+\sigma
T_{m\sigma,1\bar{\sigma}}]|_{\omega=\varepsilon_F } \label{charge}
\end{eqnarray}
obeys the Landauer-B\"{u}ttiker formula.\cite{Datta} With respect to
the spin current, it can be defined as
$J_{ms}=J_{m\uparrow}-J_{m\downarrow}$ with $J_{ms}=\mathcal
{G}_{ms}\cdot V_s$ and
\begin{eqnarray}
\mathcal
{G}_{ms}=\frac{e^{2}}{h}\sum_\sigma[T_{m\sigma,1\downarrow}-T_{m\sigma,1\uparrow}]|_{\omega=\varepsilon_F.
} \label{spin}
\end{eqnarray}
It is consequently found that in the linear regime, by only
investigating the characteristics of the linear conductances, the
properties of the spin-bias-driven charge and spin currents can be
clarified. From Eq.(\ref{charge}) and Eq.(\ref{spin}), one can
readily see that in the absence of any spin-dependent fields the
electron transmission is irrelevant to the electron spin. And then
the opposite-spin currents driven by the spin bias flow through this
ring with the same magnitude, leading to the result of zero ${\cal
G}_{mc}$ and ${\cal G}_{ms}$[ see the dashed line in
Fig.\ref{Structure}(b)].
\par
As mentioned in the recent researches\cite{Rashba3}, in some QD
structures, local Rashba interaction could efficiently modulate the
quantum interference and bring about the spin polarization in the
electron transport process. Namely, when a QD subject to Rashba
interaction is embedded in the mesoscopic interferometer, the
traveling electrons acquire a spin-dependent phase in addition to
the Aharonov-Bohm phase, which helps manipulate the electron spin
via the electric means\cite{Rashba2}. We then introduce a local
Rashba interaction to QD-2 of this structure and aim to investigate
its charge and spin properties influenced by the interplay between
the spin bias and Rashba interaction. As shown in
Fig.\ref{Structure}(b), in the presence of Rashba SO coupling and
the absence of magnetic field, there indeed emerge apparent charge
currents in the drain terminals. Moreover, an interesting phenomenon
is that in the whole regime the amplitude of $J_{2c}$ is the same as
that of $J_{3c}$ but the directions of them are always opposite to
each other. Such a result suggests that by building a closed circuit
between lead-2 and lead-3 the feature of the spin bias in lead-1 can
be measured by observing the charge current flowing between the
drain terminals. Surely, the magnitudes of charge currents are
related to the values of QD levels with respect to the energy zero
point, i.e., in the vicinity of $\varepsilon_0={\Gamma\over 2}$ the
charge currents reach the extremum of them. On the other hand, since
the configuration of quantum ring, we now would like to investigate
the effect of a local magnetic flux on the electron motion in this
system. In Fig.\ref{Structure}(c) we see that the application of
magnetic flux can further adjust the spin-bias-induced charge
currents. And, with the tuning of magnetic flux the charge currents
($J_{2c}$ and $J_{3c}$) oscillate with the period $\Delta\phi=2\pi$.
By the increase of magnetic flux from $\phi=0$ to $\phi=\pi$
$J_{2c}$ encounters its minimum whereas $J_{3c}$ arrives at its
maximum with their zero value at $\phi =\frac{1}{2}\pi$. Therefore,
the magnetic flux can effectively vary the magnitude and direction
of the charge currents.

\par
With respect to the spin current, in the linear regime, it arises
from the spin polarization in the electron transport process [see
Eq.(\ref{spin})]. And, in the absence of Rashba interaction, a small
spin bias cannot bring about the spin polarization in the drain
terminals. But when the Rashba interaction is taken into account,
there emerge notable spin currents in lead-2 and lead-3, as shown in
Fig.\ref{Structure}(d). Note that, the difference between these two
spin currents originates from the contribution of the interdot spin
flip terms in the Hamiltonian. In addition, we find that the spin
current directions can be efficiently modulated via the magnetic
flux. For the case of $\phi={\pi\over 2}$ there presents obvious
spin polarization in lead-2, but there is little spin polarization
in lead-3. Alternatively, when the magnetic flux is increased to
$\phi={3\pi\over 2}$ the opposite result comes into being, i.e., the
spin polarization in lead-3 reaches its maximum. Thereby, in such a
structure, by the cooperation of the Rashba interaction and magnetic
flux the highly-polarized spin current in either drain terminal can
be alternately obtained.
\begin{figure}
\begin{center}
\scalebox{0.4}{\includegraphics{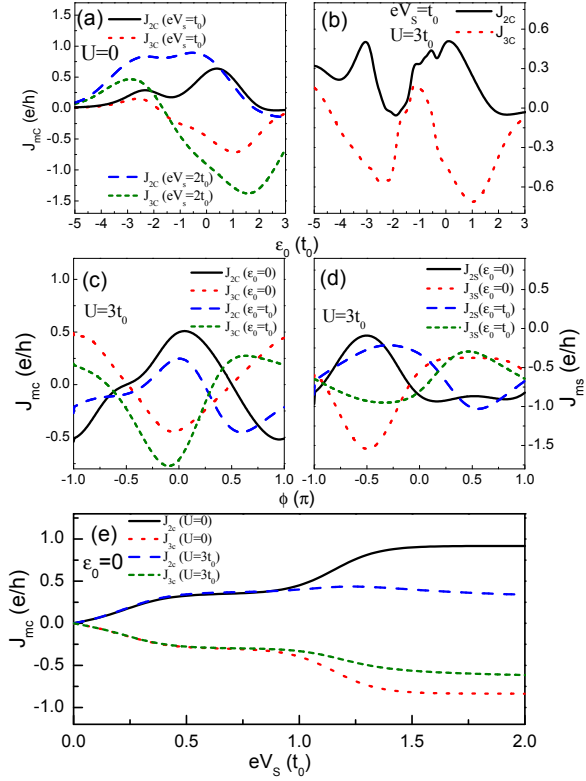}} \caption{ The current
results in the cases of the finite spin bias. The temperature is
assumed to be $k_BT=0.1t_0$. (a)-(b) The charge currents vs QD
levels with $U=0$ and $U=3t_0$, respectively. (c)-(d) The charge and
spin currents as functions of the magnetic flux. (e) The charge
currents versus the spin bias strength.\label{Fbias}}
\end{center}
\end{figure}
\par
It is evident that the above results are dependent on the electron
transport properties in this structure. Then in order to clarify
these results we focus on the transmission functions, as shown in
Fig.\ref{Trans} with $\varepsilon_j=0$. They are just the integrands
for the calculation of the charge currents [see Eq.(\ref{current})].
Here it is necessary to emphasize that although the Rashba-related
spin-flip terms contribute to the electron transport, the
spin-conserved electron motion determines the transport results of
this system.\cite{gongapl} So, to keep the argument simple, we drop
the spin flip terms for the analysis of electron transport
behaviors. By comparing the results shown in Fig.\ref{Trans}(a), we
can readily see that in the absence of magnetic flux, the traces of
$T_{2\uparrow,1\uparrow}$ and $T_{3\downarrow,1\downarrow}$ coincide
with each other very well, so do the curves of
$T_{2\downarrow,1\downarrow}$ and $T_{3\uparrow,1\uparrow}$.
Substituting this result into Eq.(\ref{charge}), one can certainly
arrive at the result of the distinct charge currents in the drain
terminals. On the other hand, these transmission functions depend
nontrivially on the magnetic phase factor, as exhibited in
Fig.\ref{Trans}(b) with $\phi={\pi\over 2}$. In comparison with the
zero magnetic field case, herein the spectra of
$T_{j\downarrow,j'\downarrow}$ are reversed about the axis
$\omega=0$ without the change of their amplitudes, but the
amplitudes of $T_{j\uparrow,j'\uparrow}$ present a thorough change.
Similarly, with the help of Eq.(\ref{charge}), one can then
understand the disappearance of charge currents in such a case.
Moreover, the presentation of spin currents can be understood with
the help of above result and Eq.(\ref{spin}).
\par
The underlying physics being responsible for the spin dependence of
the transmission functions is quantum interference, which manifests
if we analyze the electron transmission process in the language of
Feynman path. Therefore based on this method, we write
$T_{2\sigma,1\sigma}=|\tau_{2\sigma,1\sigma}|^2$ where the
transmission probability amplitude is defined as
$\tau_{2\sigma,1\sigma}=\widetilde{V}^*_{2\sigma}
G^r_{2\sigma,1\sigma}\widetilde{V}_{1\sigma}$ with
$\widetilde{V}_{j\sigma}=V_{j\sigma}\sqrt{2\pi\rho_j(\omega)}$. With
the solution of $G^r_{2\sigma,1\sigma}$, we find that the
transmission probability amplitude $\tau_{2\sigma,1\sigma}$ can be
divided into three terms, i.e.,
$\tau_{2\sigma,1\sigma}=\tau^{(1)}_{2\sigma,1\sigma}+\tau^{(2)}_{2\sigma,1\sigma}$,
where
$\tau^{(1)}_{2\sigma,1\sigma}=\frac{1}{D}\widetilde{V}^*_{2\sigma}
g_{2\sigma}t^*_{1\sigma}g_{1\sigma}\widetilde{V}_{1\sigma}$ and
$\tau^{(2)}_{2\sigma,1\sigma}=\frac{1}{D}\widetilde{V}^*_{2\sigma}
g_{2\sigma}t_{2\sigma}g_{3\sigma}t_3e^{i\phi}
g_{1\sigma}\widetilde{V}_{1\sigma}$ with
$D=\det\{[G^r]^{-1}\}\prod_jg_{j\sigma}$. By observing the
structures of $\tau^{(1)}_{2\sigma,1\sigma} $ and
$\tau^{(2)}_{2\sigma,1\sigma} $, we can readily find that they just
represent the two paths from lead-2 to lead-1 via the QD ring. The
phase difference between $\tau^{(1)}_{2\sigma,1\sigma}$ and
$\tau^{(2)}_{2\sigma,1\sigma}$ is
$\Delta\phi_{2\sigma}=[\phi-2\sigma\varphi+\theta_3]$ with
$\theta_j$ arising from $g_{j\sigma}$. It is clearly known that only
such a phase difference is related to the spin polarization.
$T_{3\sigma,1\sigma}$ can be analyzed in a similar way. We then
write
$T_{3\sigma,1\sigma}=|\tau^{(1)}_{3\sigma,1\sigma}+\tau^{(2)}_{3\sigma,1\sigma}|^2$,
with $\tau^{(1)}_{3\sigma,1\sigma}
=\frac{1}{D}\widetilde{V}^*_{3\sigma}g_{3\sigma}t_{3}e^{i\phi}g_{1\sigma}\widetilde{V}_{1\sigma}
$ and
$\tau^{(2)}_{3\sigma,1\sigma}=\frac{1}{D}\widetilde{V}^*_{3\sigma}g_{3\sigma}t^*_{2\sigma}g_{2\sigma}t^*_{1\sigma}g_{1\sigma}\widetilde{V}_{1\sigma}
$. The phase difference between $\tau^{(1)}_{4\sigma,1\sigma}$ and
$\tau^{(2)}_{3\sigma,1\sigma}$ is
$\Delta\phi_{3\sigma}=[\phi-2\sigma\varphi-\theta_2]$. Utilizing the
parameter values in Fig.\ref{Trans}, we evaluate that
$\varphi\approx{\pi\over 7}$ and $\theta_j=-\frac{\pi}{2}$ at the
point of $\omega=0$. It is apparent that when $\phi=0$ only the
phase differences $\Delta\phi_{2\sigma}$ are spin-dependent.
Accordingly, we obtain that $\Delta\phi_{2\uparrow}={-11\pi\over
14}$, $\Delta\phi_{2\downarrow}={3\pi\over 14}$,
$\Delta\phi_{3\uparrow}={3\pi\over 14}$, and
$\Delta\phi_{3\downarrow}={11\pi\over 14}$, which clearly prove that
the quantum interference between $\tau^{(1)}_{2\uparrow,1\uparrow}$
and $\tau^{(2)}_{2\uparrow,1\uparrow}$
($\tau^{(1)}_{3\downarrow,1\downarrow}$ and
$\tau^{(2)}_{3\downarrow,1\downarrow}$ alike) is destructive, but
the constructive quantum interference occurs between
$\tau^{(1)}_{2\downarrow,1\downarrow}$ and
$\tau^{(2)}_{2\downarrow,1\downarrow}$
($\tau^{(1)}_{3\uparrow,1\uparrow}$ and
$\tau^{(2)}_{3\uparrow,1\uparrow}$ alike). Then such a quantum
interference pattern can explain the traces of the transmission
functions shown in Fig.\ref{Trans}(a). In the case of
$\phi=\frac{\pi}{2}$ we find that only $\Delta\phi_{2(3)\sigma}$ are
crucial for the occurrence of spin polarization. By a calculation,
we obtain $\Delta\phi_{2\uparrow}=-{2\pi\over 7}$,
$\Delta\phi_{2\downarrow}={2\pi\over 7}$,
$\Delta\phi_{3\uparrow}={5\pi\over 7}$, and
$\Delta\phi_{3\downarrow}={9\pi\over 7}$, which are able to help us
clarify the results in Fig.\ref{Trans}(c) and (d). Up to now, the
characteristics of the transmission functions, as shown in
Fig.\ref{Trans}, hence, the tunability of charge currents have been
clearly explained by analyzing the quantum interference between the
transmission paths.
\par
For the case of finite spin bias, the charge currents in lead-2 and
lead-3 can be evaluated by Eq.(\ref{current}). Accordingly, in
Fig.\ref{Fbias}(a)-(b) we plot the charge current spectra \emph{vs}
the QD levels. In Fig.\ref{Fbias}(a), we find that different from
the linear transport results, the current spectra exhibit
complicated properties with the shift of QD levels, since the
currents are not proportional to the transmission function any more.
For the case of $eV_s=t_0$, the quantitative relation between these
two charge currents ( i.e., $J_{2c}=-J_{3c}$) is seeable ( only in
the region of $\varepsilon_0$ greater than $-t_0$ and less than
$t_0$ ). When the spin-bias strength is increased to $eV_s=2t_0$,
the magnitude of the charge currents increase. Meanwhile, it is seen
that the curve of $J_{2c}$ tends to be symmetric and the profile of
$J_{3c}$ becomes asymmetric. Consequently, here the relation of
$J_{2c}=-J_{3c}$ becomes ambiguous except at the position of
$\varepsilon_0=0$.
\par
By far, we have not discussed the effect of electron interaction on
the occurrence of charge currents in the drain terminals, though it
is included in our theoretical treatment. Now we incorporate the
electron interaction into the calculation with $U_j=U=3t_0$, and we
deal with the many-body terms by employing the second-order
approximation, since such an approximation is feasible for the case
the system temperature higher than the Kondo temperature (where the
electron correlation is comparatively weak). Fig.\ref{Fbias}(b)
shows the calculated currents spectra vs the QD levels with
$k_BT=0.1t_0$. From this figure, we see that the many-body effect
causes the further oscillation of the charge current spectra. It is
obvious that the intradot electron interaction splits the current
curves into two groups, and in each group the current properties are
analogous to those in the noninteracting case. This is because that
within such an approximation the Coulomb interaction only gives rise
to the splitting of the QD level, i.e., $\varepsilon_j$ and
$\varepsilon_j+U$. As a result, one can find in such a case, in the
high-energy and low-energy regimes (i.e., \emph{around the points of
$\varepsilon_0=0$ and $\varepsilon_0=-3t_0$}) the result of
$J_{2c}=-J_{3c}$ is also seeable. On the other hand, it shows that
the effect of the Coulomb interaction on the spin currents is
nontrivial, and in the region of $0<\phi<\pi$ the magnitudes of the
spin currents are suppressed [see Fig.\ref{Fbias}(d)].
\par
Finally, we investigate the charge currents as functions of the
spin-bias strength, with the calculated results shown in
Fig.\ref{Fbias}(e). First, at the noninteracting case, we see that
in the situation of $0<eV_s<{t_0\over 2}$ (or $t_0<eV_s<{3t_0\over
2}$), the charge currents increase with the strengthening of spin
bias. But next when the spin bias strength varies in the regime of
${t_0\over 2}<eV_s<t_0$ (or ${3t_0\over 2}<eV_s<2t_0$), the
magnitudes of the charge currents change a little. However, when the
many-body effect is taken into account, it is found that only in the
situation of $0<eV_s<{t_0\over 2}$, the magnitudes of the charge
currents are proportional to the strengthening of spin bias, whereas
in the cases the charge currents are approximately independent of
the variation of the spin bias.

\section{Summary}
\par
In summary, in the present triple-QD ring, the local Rashba
interaction provides a spin-dependent AB phase difference. The
three-terminal configuration balances the electron transmission
probabilities via two different arms of the QD ring. The variation
of the magnetic field strength and the QD level can adjust the phase
difference between the two kinds of Feynman paths on an equal
footing. Thus, the spin dependence of the electron transmission
probability can be controlled by altering the exerted magnetic field
or the QD levels. Then with the consideration of a spin bias in one
terminal, it is possible to obtain the tunable charge and spin
currents in the either two terminals. On the other hand, we readily
emphasize that when the spin bias is considered in another
terminal(e.g., lead-2), the direction of the charge current in
lead-3 will be inverted since the geometry of this structure.
However, if the Rashba interaction is applied in another QD (e.g.,
QD-2) we can also see the inversion of the current directions in
lead-3. That is to say, the current directions in either drain
terminal can be modulated by controlling the spin bias or Rashba
interaction. So, such a structure can be proposed to be a prototype
of a charge and spin current rectifier.
\par

\section*{Acknowledgments}

This work was financially supported by the National Natural Science
Foundation of China (Grant No. 10904010), the Seed Foundation of
Northeastern University of China (Grant No. N090405015), and the
Scientific Research Project of Liaoning Education Office (Grant No.
2009A309).

\clearpage

\bigskip


\begin{thebibliography}{99}
\bibitem{Ohno} Y. Ohno, D. K. Young, B. Beschoten, $et$ $al.$,
Nature, 402 (1999) 790.


\bibitem{Dasarma} Zutic, J. Fabian, and S. Das Sarma, Rev. Mod. Phys. 76
(2004) 323; R. Hanson, L. P. Kouwenhoven, J. R. Petta, S. Tarucha,
and L. M. K. Vandersypen, Rev. Mod. Phys. 79 (2004) 1217; S. A.
Wolf, $et$ $al$., Science 294 (2001) 1488.

\bibitem{Loss} D. Loss, and D. P. DiVincenzo, Phys. Rev. A 57 (1998) 120; G.
Burkard, D. Loss, and D. P. DiVincenzo, Phys. Rev. B 63 (1999) 2070.
\bibitem{Loss2} D. V. Bulaev and D. Loss, Phys. Rev. B 71
(2005) 205324.
\bibitem{Datta} A. A. Kiselev and K. W. Kim, Appl. Phys. Lett. 78 (2001) 775;
T. P. Pareek, Phys. Rev. Lett. 92 (2004) 076601.

\bibitem{Rashba} J. Nitta,
T. Akazaki, H. Takayanagi, and T. Enoki, Phys. Rev. Lett. 78 (1997)
1335.

\bibitem{Rashba2} G. Engels, J. Lange, Th. Sch\"{a}pers, and H. L\"{u}th, Phys. Rev. B
55, R1958 (1997); D. Grundler, Phys. Rev. Lett. 84 (2000) 6074.
\bibitem{Sun} Q. F. Sun, J. Wang, and H. Guo, Phys. Rev. B 71
(2005) 165310; B. K. Nikolic and S. Souma, Phys. Rev. B 71 (2005)
195328.
\bibitem{Trocha} P. Trocha and J. Barnas, Phys. Rev. B 76
(2007) 165432; P. Trocha and J. Barnas, Phys.: Condens. Matter 20
(2008) 125220.

\bibitem{Other} M. L. Ladr\'{o}n de Guevara and P. A. Orellana , Phys. Rev. B 73,
(2006) 205303.

\bibitem{Nagaosa} A. Brataas, Y. Tserkovnyak, G. E. W. Bauer and B. I. Halperin, Phys. Rev. B 66 (2002) 060404;
S. O. Valenzuela and M. Tinkham, Nature (London) 442 (2006) 176; Y.
K. Kato, R. C. Myers, A. C. Gossard and D. D. Awschalom, Science 306
(2004) 1910; S. D. Ganichev, \emph{et al.}, Nature Physics 2 (2006)
609.

\bibitem{Hubner} J. Hubner, et al., Phys. Rev. Lett., 90 (2003) 216601; Martin J.
Stevens \emph{et al.}, Phys. Rev. Lett, 90 (2003) 136603.
\bibitem{Cui} M. Y. Veillette, C. Bena, and L. Balents, Phys. Rev. B 69 (2004)
075319; X.-D. Cui, S.-Q. Shen, J. Li, Y. Ji, W. Ge, and F.-C. Zhang,
Appl. Phys. Lett. 90 (2007) 242115.
\bibitem{Li} J. Li and S. Q. Shen, Phys. Rev. B 76 (2007) 153302; P. Zhang, Q. K. Xue, and
X. C. Xie, Phys. Rev. Lett. 91 (2003) 196602.
\bibitem{Jpn} H. Katsura,J. Phys. Soc. Jpn. 76
(2007) 054710; Y.-J. Bao, N.-H. Tong, Q.-F. Sun, and S.-Q. Shen,
Europhys. Lett. 83 (2008) 37007.
\bibitem{LuHZ} H.-Z. Lu and S.-Q. Shen, Phys. Rev. B 77
(2008) 235309; H.-Z. Lu, B. Zhou, and S.-Q. Shen, Phys. Rev. B 79
(2009) 174419; H.-Z. Lu and S.-Q. Shen, Phys. Rev. B 80 (2009)
094401.
\bibitem{Chi} F. Chi, and X. Q. Yuan, Chi. Phys. Lett. 26,(2009)
097301.

\bibitem{Bao} K. Bao and Y. Zheng, Phys. Rev. B 73 (2006) 045316.
\bibitem{Tarucha} S. Amaha, T. Hatano, T. Kubo, S. Teraoka, Y. Tokura, S. Tarucha, and
D. G. Austing, Appl. Phys. Lett. 94 (2009) 092103.
\bibitem{gongapl} W. Gong, Y. Zheng, and T. L\"{u}, Appl. Phys. Lett. \textbf{92},
042104 (2008); W. Gong, Y. Han, and G. Wei, Sol. Stat. Comm.
\textbf{149}, 1831(2009).
\bibitem{Serra} D. S\`{a}nchez and L. Serra, Phys. Rev. B \textbf{74}, 153313
(2006).
\bibitem{Meir} Y. Meir and N. S. Wingreen, Phys. Rev. Lett. \textbf{68},
2512 (1992); A.-P. Jauho, N. S. Wingreen, and Y. Meir, Phys. Rev. B
\textbf{50}, 5528 (1994).
\bibitem{Gong1} W. Gong, Y. Zheng, Y. Liu, and T.
L\"{u}, Phys. Rev. B 73 (2006) 245329.


\bibitem{ZhengHZ} J. Q. You and H. Z. Zheng, Phys. Rev. B 60 (1999) 13314;
J. Q. You and H. Z. Zheng, Phys. Rev. B 60 (1999) 8727.





\bibitem{Sarra} J. Nitta, T. Akazaki, H. Takayanagi, and T. Enoki, Phys. Rev. Lett. \textbf{78},
1335 (1997); F. Mireles and G. Kirczenow, Phys. Rev. B \textbf{64},
024426 (2001).

\bibitem{Rashba3} A. M. Lobos and A. A. Aligia, Phys. Rev. Lett.
\textbf{100}, 016803 (2008); R. J. Heary, J. E. Han, and Lingyin Zhu
Phys. Rev. B 77, 115132 (2008); R. Citro, and F. Romeo, Phys. Rev. B
\textbf{77}, 193309 (2008).
\end{thebibliography}
\end{document}